\documentclass[fleqn,usenatbib]{mnras}

\usepackage{newtxtext,newtxmath}

\usepackage[T1]{fontenc}

\DeclareRobustCommand{\VAN}[3]{#2}
\let\VANthebibliography\thebibliography
\def\thebibliography{\DeclareRobustCommand{\VAN}[3]{##3}\VANthebibliography}

\usepackage{graphicx}	
\usepackage{amsmath}	

\title[PEARLS: NEP High-z Ring]{JWSTs PEARLS: NIRCam imaging and NIRISS spectroscopy of a $z=3.6$ star-forming galaxy lensed into a near-Einstein Ring by a $z=1.258$ massive elliptical galaxy}

\author[N. J. Adams et al.]{
Nathan J. Adams,$^{1}$\thanks{E-mail: nathan.adams@manchester.ac.uk}
Giovanni Ferrami,$^{2,3}$
Lewi Westcott,$^{1}$
Thomas Harvey,$^{1}$
Vicente Estrada-Carpenter,$^{4}$ \newauthor
Christopher J. Conselice,$^{1}$
Duncan Austin,$^{1}$
J. Stuart B. Wyithe,$^{2,3,5}$
Caio M. Goolsby,$^{1}$
Qiong Li,$^{1}$ \newauthor
Vadim Rusakov,$^{1}$ 
Rogier A. Windhorst,$^{4}$
Seth H. Cohen,$^{4}$
Rolf A. Jansen,$^{4}$
Jake Summers,$^{4}$\newauthor
Roselia O'Brein,$^{4}$
Anton M. Koekemoer,$^{6}$ 
Simon P. Driver,$^{7}$
Brenda Frye,$^{8}$
Nimish P. Hathi,$^{6}$
Dan Coe,$^{6,9,10}$\newauthor
Norman A. Grogin,$^{6}$
Madeline A. Marshall,$^{11}$
Nor Pirzkal,$^{6}$
Russell E. Ryan Jr.,$^{6}$\newauthor
Christopher N. A. Willmer,$^{8}$
Haojing Yan,$^{12}$
Benne W. Holwerda,$^{13}$
Patrick S. Kamieneski,$^{4}$\newauthor
Tom Broadhurst,$^{14,15,16}$
W. Peter Maksym,$^{17}$
Payaswini Saikia,$^{18}$
Joseph D. Gelfand$^{18}$
\\
$^{1}$Jodrell Bank Centre for Astrophysics, University of Manchester, Oxford Road, Manchester M13 9PL, UK\\
$^{2}$School of Physics, University of Melbourne, Parkville, VIC 3010, Australia\\
$^{3}$ARC Centre of Excellence for All-Sky Astrophysics in 3 Dimensions (ASTRO 3D)\\
$^{4}$School of Earth and Space Exploration, Arizona State University, Tempe, AZ 85287-1404, USA\\
$^{5}$Research School of Astronomy and Astrophysics, Australian National University, Canberra, ACT 2611, Australia\\
$^{6}$Space Telescope Science Institute,
3700 San Martin Drive, Baltimore, MD 21218, USA\\
$^{7}$International Centre for Radio Astronomy Research (ICRAR) and the
International Space Centre (ISC), The University of Western Australia,\\
M468, 35 Stirling Highway, Crawley, WA 6009, Australia\\
$^{8}$Department of Astronomy/Steward Observatory, University of Arizona, 933 N Cherry Ave,
Tucson, AZ, 85721-0009, USA\\
$^{9}$Association of Universities for Research in Astronomy (AURA) for the European Space Agency (ESA), STScI, Baltimore, MD 21218, USA\\
$^{10}$Center for Astrophysical Sciences, Department of Physics and Astronomy, The Johns Hopkins University, 3400 N Charles St. Baltimore, MD 21218, USA\\
$^{11}$Los Alamos National Laboratory, Los Alamos, NM 87545, USA\\
$^{12}$Department of Physics and Astronomy, University of Missouri,
Columbia, MO 65211, USA\\
$^{13}$University of Louisville, Department of Physics and Astronomy, 102 Natural Science Building, 40292 KY Louisville, USA\\
$^{14}$Donostia International Physics Center, DIPC, Basque Country, San Sebastián, 20018, Spain\\
$^{15}$Department of Physics, University of Basque Country UPV/EHU, Bilbao, Spain\\
$^{16}$Ikerbasque, Basque Foundation for Science, Bilbao, Spain\\
$^{17}$NASA Marshall Space Flight Center, Huntsville, AL 35812, USA \\
$^{18}$Center for Astrophysics and Space Science, New York University Abu Dhabi, PO Box 129188, Abu Dhabi, UAE\\
}

\date{Accepted XXX. Received YYY; in original form ZZZ}

\pubyear{\the\year{}}

\begin{document}
\label{firstpage}
\pagerange{\pageref{firstpage}--\pageref{lastpage}}
\maketitle

\begin{abstract}
We present the discovery, and initial lensing analysis, of a high-redshift galaxy-galaxy lensing system within the JWST-PEARLS/HST-TREASUREHUNT North Ecliptic Pole Time Domain Field (designated NEPJ172238.9+655143.1). The lensing geometry shears a $z=3.6\pm0.1$ star-forming galaxy into a near-Einstein ring with a radius of 0\farcs92, consisting of 4 primary images, around a foreground massive elliptical galaxy at $z=1.258\pm0.005$. The system is fortuitously located within the NIRISS F200W footprint of the PEARLS survey, enabling spectroscopic identification of the 8500{\AA} TiO band in the foreground galaxy and allowing tight constraints to be placed on the redshift of the background galaxy based on its continuum detection and lack of strong emission lines. We calculate magnification factors of $2.6<\mu<8.4$ for the four images and a total lensing mass of $(4.08 \pm 0.07)\times10^{11}M_\odot$. SED fitting of the foreground elliptical galaxy within the Einstein radius reveals a stellar mass of $\sim1.26\times10^{11}M_\odot$, providing a mass/light ratio of 3.24. Employing simple scaling relations and assumptions, an NFW dark matter halo is found to provide the correct remaining mass within $0.12^{+0.21}_{-0.09}$dex. However, if a bottom-heavy IMF for elliptical galaxies is employed, stellar mass estimations increase and can account for the majority of the lensing mass (up to $\sim$83\%), reducing the need for dark matter. This system further demonstrates the new discovery space that the combined wavelength coverage, sensitivity and resolution of JWST now enables.
\end{abstract}

\begin{keywords}
gravitational lensing: strong -- galaxies: elliptical and lenticular, cD -- stars: luminosity function, mass function -- cosmology: dark matter
\end{keywords}



\section{Introduction} \label{sec:intro}

Gravitational lensing is a phenomena that astronomers have been exploiting to enhance our understanding of the Universe for decades now. Applications vary from magnifying the faintest high-redshift galaxies with the aid of massive galaxy clusters \citep[recent examples include][]{Atek2023,Bradley2023,Ma2024,Fujimoto2024} to measuring the expansion rate of the Universe \citep[e.g.][]{Pierel2024,Frye2024,Pascale2025}.
In rare cases, lensing from individual high-mass galaxies can occur. Galaxy-galaxy lensing provides opportunities to not only exploit the benefits of magnification on background sources, but also probes mass distributions within individual galaxies beyond what can be measured from their stellar emission. In rarer instances still, the geometry of a lensing system can result in an Einstein ring, where the light from a background galaxy is stretched into a near continuous ring-shaped image \citep{Hewitt1988,Jauncey1991,King1998}.  This is a rare case in the lensing configuration where the source is directly behind the lens.  Such images have been observed from a selection of surveys and compilation programmes using both ground-based and space-based observatories \citep[e.g.][]{More2012,Sonnenfield2020,Garvin2022,Nightingale2025,Riordan2025,Walmsley2025}. 

The geometry of an Einstein ring system enables the determination of the total mass contained within the rings radius at the distance of the lens. When compared to the stellar mass estimated within the same radius (from e.g. SED modelling) and dynamical masses from resolved spectroscopy, the contributions from luminous and non-luminous matter can be directly compared \citep[e.g.][]{Barnabe2011,Thomas2011}. Strong galaxy-galaxy lensing can subsequently be used to explore total projected mass distributions within $\sim1$" in massive, elliptical galaxies and provide a testing ground for models of dark matter haloes \citep[e.g.][]{Mercier2024}.

With the advent of the James Webb Space Telescope (JWST), the combined wavelength range, sensitivity and resolution within the near-infrared has opened up new parameter space to identify and study such Einstein rings and galaxy-galaxy lensing systems in general \citep{Holloway2023,Ferrami2024,Hogg2025}. Ultimately, these systems are expected to be identified at higher and higher redshifts for both the foreground lens and background galaxy being lensed \citep[though JWST has also proven capable of using very low-z galaxies as lenses, e.g.][]{Keel2023}. JWST has already confirmed at least one Einstein ring system within the COSMOS-Web field with a high potential lens redshift of $z\sim2$ and a background galaxy at $z\sim5.1$ \citep{VanDokkum2023,Mercier2024,Shuntov2025}. In light of this finding, larger-scale searches for galaxy-galaxy lensing within JWST's growing archival dataset is already underway (Ferrami et al. In Prep). The first large-scale search, using the completed COSMOS-Web survey, has revealed over 100 instances of strong galaxy-galaxy lensing \citep{Nightingale2025,Mahler2025}. JWST is also not the only facility pushing this boundary, \emph{Euclid's} Q1 release identified some 497 candidate instances of galaxy-galaxy lensing which has doubled the total number with space-based imaging \citep{Walmsley2025}. Identifying significant populations of higher-z lensing systems are of prime interest since their occurrence rates and distributions are dependent upon other population-based observables that are widely studied in extragalactic astronomy, such as the distributions of lens mass profiles \citep[e.g. single isothermal ellipsoids][]{Kormann1994, TreuKoopmans2002}, UV luminosity function \citep[e.g.][]{Moutard2020,Bouwens2021, Adams2024, FerramiUVLF2023}, luminosity-size relation \citep[e.g.][]{Shibuya2015}, velocity dispersion function \citep[see discussion in e.g.][]{Treu2005,Belli2014,Mason2015,FerramiVDF2025}, ellipticity distribution \citep{vanderWel2014} and the fundamental plane \citep[FP: relationship between surface brightness, velocity dispersion and effective radius e.g.][]{Treu2001}.

In this paper, we present a newly identified candidate Einstein ring system located within NIRCam imaging and NIRISS grism spectroscopy of the North Ecliptic Pole Time Domain Field \citep[NEP-TDF][]{Jansen2018}, the primary blank-field component of JWST's multi-field Prime Extragalactic Areas for Reionization and Lensing Science (PEARLS) survey \citep{Windhorst2023}. This system consists of a high redshift lens, spectroscopically confirmed at $z=1.258\pm0.005$, which magnifies a higher redshift background source in a galaxy-galaxy lensing system at $z=3.6\pm0.1$.   We discuss the discovery of this unusual system, including the properties of the foreground and background galaxies as well as the mass of the foreground dark matter halo.  

Below we give a short description of our paper's outline.  In Section 2 we discuss the data used and their reduction processes. In Section 3 we model the photometry, grism spectra and lensing properties of the system. Finally, in section 4 we discuss implications for dark matter,the initial mass function of ellipticals and the chances of observing such a system within archival JWST datasets before summarising our conclusions in Section 5. All calculations involving a cosmological model assume $\Lambda$CDM with $H_0=70$\,km\,s$^{-1}$\,Mpc$^{-1}$, $\Omega_{\rm M}=0.3$ and $\Omega_{\Lambda} = 0.7$. All magnitudes listed follow the AB magnitude system \citep{Oke1974,Oke1983}.

\section{Data}

\subsection{JWST Imaging}

The system was first identified by eye within NIRCam imaging of the NEP-TDF field taken during the 3rd visit of the PEARLS GTO programme \citep[PID: 2738, PI: R. Windhorst][]{Windhorst2023}. This field is located within JWSTs continuous viewing zone, meaning it is accessible at all times of the year. The PEARLS data consists of 8 partially overlapping NIRCam pointings grouped into 4 pairs, with each pointing pair separated by 3 months. This produces a cross/windmill shaped geometry for the field. NIRCam data were reduced as part of the EPOCHS project, using JWST pipeline version 1.8.2, calibration pmap1084 and a pixel scale of 0\farcs03 pix$^{-1}$ \citep{Adams2024,Conselice2024}. In addition to standard calibrations, we employ the use of a 1/f correction developed by Chris Willott \footnote{\url{https://github.com/chriswillott/jwst/tree/master}} and subtract scaled templates of `Wisp' artefacts in the F150W and F200W images. The 8 NIRCam bands (F090W, F115W, F150W, F200W, F277W, F356W, F410M, and F444W) cover a total area of $\sim$60 arcmin$^2$. We calculate $5\sigma$ point-source depths ranging from 28.5-29.3 within 0.\farcs32 diameter apertures ($\sim80\%$ enclosed PSF flux) from the random scattering of empty apertures across the final mosaic.

\begin{figure*}
    \centering
    \includegraphics[width=0.8\textwidth]{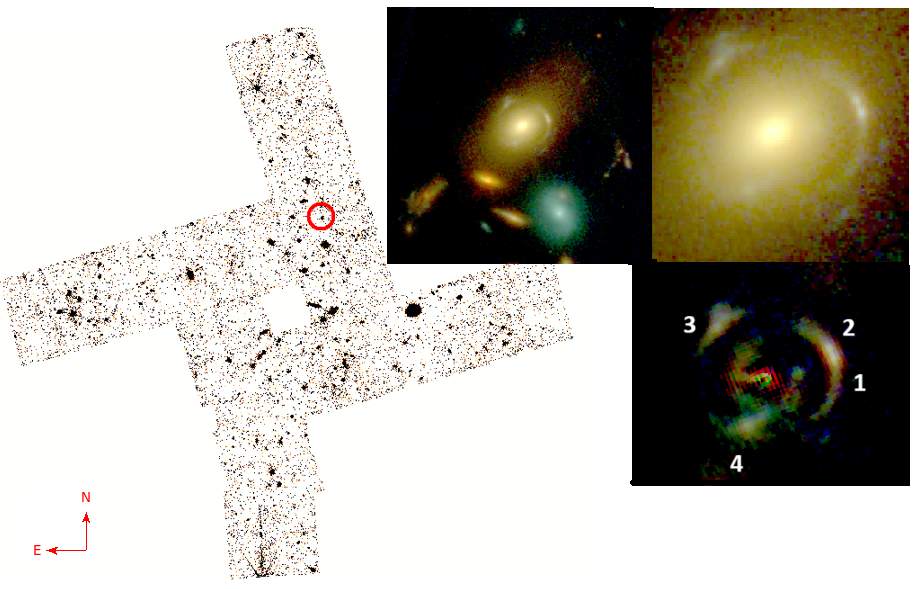}
    \caption{The full F200W NIRCam footprint of the NEP-TDF Field with the object of interested highlighted with a red circle. Accompanying RGB cutouts use the F444W, F200W, F090W filters and show: {\bf Top Left} a $\sim10\times10$ arcsecond cutout of the region surrounding the target of interest. {\bf Top Right} a zoom into the core of the elliptical highlighting the ring-like structure which has a radius of 0\farcs92. {\bf Bottom Right} A zoom into the residual lensed arcs once a Sérsic  profile for the elliptical galaxy is subtracted off and the arcs numbered.}
    \label{fig:Field}
\end{figure*}

\subsection{HST Imaging}

To aid photometric analysis, we include HST imaging in the F606W filter from the TREASUREHUNT programme \citep[][Jansen et al. In prep, GO 15278, PI: R. Jansen and GO 16252/16793, PIs: R. Jansen \& N. Grogin]{Obrien2024} and is pixel matched to the JWST imaging. Data spans from Oct 2017 to Oct 2022 and provides F275W, F435W and F606W imaging for up to 194 square arcminutes. The F606W mosaic has a comparable depth to the JWST data, with a limiting $5\sigma$ magnitude of 28.5 in the same aperture size of 0.\farcs32, whilst the bluer bands are upwards of 1 magnitude shallower. For this reason, we only use the F606W mosaic in this study. Details of reduction, as well as the general field/survey geometry, are provided in \citet{Obrien2024}.

\subsection{NIRISS F200W Grism Spectroscopy}

In parallel to the NIRCam observations, NIRISS slitless spectroscopy was obtained using the orthogonal GR150R and GR150C grisms crossed with the F200W filter, yielding spectra over the 1.75--2.23\micron\ wavelength range at an approximate resolving power of $R = 220$. F200W direct images bracketed each dispersed observation. The survey design enabled the eight NIRISS pointings, with a combined area of $\sim$37 arcmin$^2$, to almost entirely overlap the central region of the NIRCam coverage of the NEP-TDF field, with exposure times similar to the longest NIRCam observations ($\sim$2800\,s) in each of the orthogonal grisms. The grism data were processed using the grism modelling and analysis software {\tt grizli} which performs an end-to-end reduction of the data and contamination modelling \citep{grizli}. An additional contamination modelling was performed using the spatially resolved methodology described in \citet{Estrada-Carpenter2024}, where 2nd, 1st, 0th, and -1st orders of the spectra were modelled.

\section{Understanding and Modelling the System}

The system was identified as a candidate Einstein ring system at the coordinates of RA = 17:22:38.98, DEC = +65:51:43.16 in the third, Northern tile of the JWST imaging of the NEP-TDF field (See Figure \ref{fig:Field}). It has been designated NEPJ172238.9+655143.1 and also NEP-TDF-L2 within the compilation of JWST identified galaxy-galaxy lenses in JWST GTO data by [Ferrami et al. In Prep].  Below we give a detailed description of our analysis of this system, including how we measured their redshifts and carried out SED and lens modelling.  

\begin{figure*}
    \centering
    \includegraphics[width=0.8\textwidth]{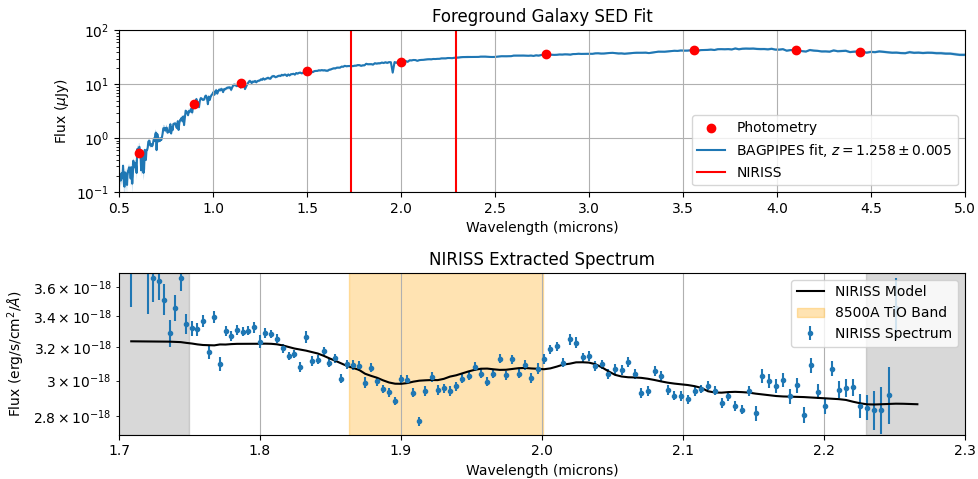} 
    \includegraphics[width=0.8\textwidth]{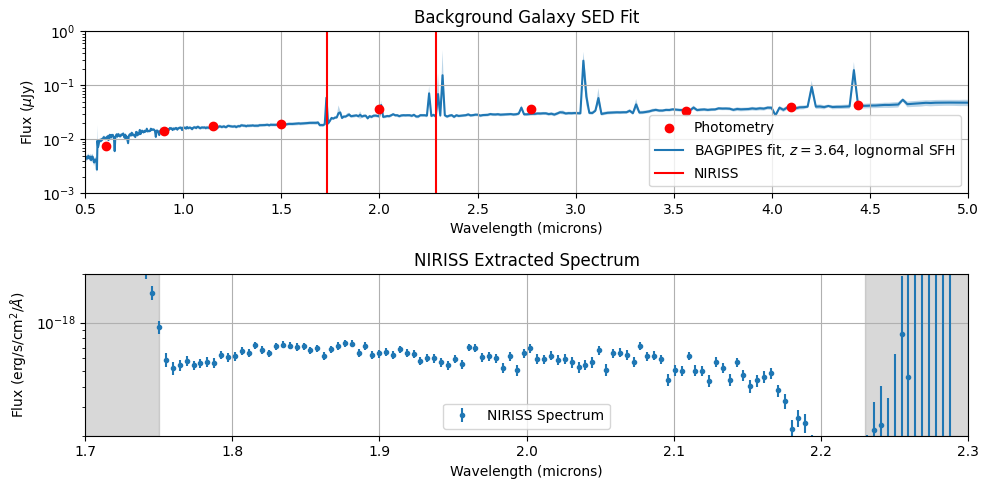}
    \caption{The best fitting SED to the extracted photometry of the foreground elliptical galaxy ($<0.9$as, Top) and the demagnified lens source (Second from Bottom). Alongside the SED fits are the NIRISS spectra within the F200W bandpass (approx. 1.75 to 2.25 microns). For the foreground elliptical, we show the forward-modelled NIRISS spectrum using the methodology of \citet{Estrada-Carpenter2024} and highlight the broad absorption feature attributed to the 8500\AA\, TiO band. The NIRISS spectrum of the background source has no significant features within the high-throughput region of the F200W bandpass (1.75-2.23 microns, low throughput shaded grey).}
    \label{fig:SEDfit}
\end{figure*}

\subsection{Modelling the Lens Galaxy}

\subsubsection{The Morphology of the Elliptical Galaxy}

We begin our analysis of this strong lensing system by morphologically modelling the foreground elliptical galaxy in order to obtain reliable photometry for both the foreground and background sources. We conduct this process by providing PSF homogenised images to the {\tt galfit} code \citep{Peng2010}. PSF homogenisation was conducted with the use of {\tt pypher} \citep{boucaud2016} with default regularisation applied. Model PSFs are generated following an empirical approach that stacks isolated stars within JWST and HST imaging \citep{Skelton2014,Whitaker2019,Weaver2024}. The HST imaging consists of a complex mosaic constructed at different position angles over a large period of time. This makes modelling the PSF more complicated as it is built up from a combination of different frames. We use the underlying image weight maps to identify and cut out a large region around our source which uses the same combination of mosaiced images to produce our empirical PSF model. 

We run {\tt galfit} following the work in \citet{Westcott2024}. Here, single and double Sérsic  profiles are fit to the elliptical galaxy with surrounding sources masked in each photometric band. Using this method we find that a single Sérsic  profile fits the source well (see Figure \ref{fig:Field} for before and after subtraction images). In the red, the source has a best-fit Sérsic  index of $n=3.43$, effective radius of 0.52as, axis ratio of 0.68 and total F444W magnitude of 19.5. In the blue, the source has a best-fit Sérsic  index of $n=2.83$, effective radius of 0.52as, axis ratio of 0.68 and total F150W magnitude of 20.4.

\subsubsection{Modelling the SED and Redshift of the Elliptical Galaxy}

To obtain an initial photometric redshift (photo-z) for the elliptical galaxy, we provide the SED fitting code EAZY-py \citep{Brammer2008} with aperture photometry extracted from within the Einstein radius using the PSF homogenised images. For the process of obtaining an initial photo-z estimate, photometric errors are fixed to 10\%. After running EAZY using the base set of FSPS models, we obtain a redshift of $z_l=1.25\pm0.14$.

The large, resolved size and luminosity results in self-contamination challenges for extracting NIRISS grism spectroscopy for this object. We process the NIRISS data following \citet{Estrada-Carpenter2024} and obtain a redshift of $z=1.258\pm0.005$, matching the peak of the initial photo-z probability density function. The key observable feature that is fit is the broad TiO band at rest-frame 8500{\AA} \citep{Allard2000,Almeida2012} which we observed at $\sim$1.9 microns (see Figure \ref{fig:SEDfit}.

Finally, we conduct a spectro-photometric fit of the elliptical galaxy using {\tt BAGPIPES} \citep{Carnall2018,Carnall2019}. To conduct this process, we renormalise the grism spectra to make the integrated F200W flux consistent between the photometry and spectroscopy. Our fiducial {\tt BAGPIPES} set-up follows the LogNormal star formation history and Kroupa initial mass function \citep{Kroupa2001} with priors as used in \citet{Harvey2025}. The elliptical galaxy is found to be very massive and passive, exhibiting $U-V$, $V-J$ colours of [1.87, 1.15] which place the source firmly in the quiescent region of the UVJ diagram \citep[e.g.][]{Williams2009}. We repeat this process twice, using the full photometry of the galaxy as measured from the morphological modelling as well as the photometry within the Einstein radius of 0\farcs92. We present the results from the final {\tt BAGPIPES} fits of our sources in Table \ref{tab:SEDtab} and in Figure \ref{fig:SEDfit}. 

\begin{table*}
\caption{Results of the SED fitting process to the total emission from the foreground elliptical galaxy (determined from Sérsic  fit), the emission within the Einstein radius of the elliptical galaxy, and the demagnified properties of the background magnified galaxy. $z_{\rm phot}$ is the original EAZY fit to photometry only, $z_{\rm Phot+Spec}$ is the improved redshift measurement when NIRISS data is folded in. Physical properties are measured using BAGPIPES with a LogNormal SFH, redshift fixed to $z=1.258$ for the elliptical and a prior of $z=3.6\pm0.1$ set for the background galaxy. Demagnified photometry from the lensing model are used for the background galaxy, the magnification factor ($\mu$) of the 4 lensed images are also provided.}
\centering
\begin{tabular}{l|lll}
     & Elliptical (Total) & Elliptical (\textless{}0\farcs92) & Background Galaxy \\ \hline
$z_{\rm phot}$    & $1.25\pm0.14$      & $1.25\pm0.14$                         & $3.74^{+0.49}_{-0.22}$               \\
$z_{\rm Phot+Spec}$    & $1.258\pm0.005$      & $1.258\pm0.005$                         & $3.60\pm0.10$               \\
$\mu$ & --- & --- & 2.62, 5.04, 8.37, 8.43  \\
Stellar Mass [$\log_{10}(M_\odot)$] &   $11.48^{+0.06}_{-0.15}$         &  $11.11^{+0.07}_{-0.08}$                             & $8.17^{+0.18}_{-0.23}$                  \\
SFR$_{\rm 10Myr}$ [$M_\odot/{\rm yr}$]  &   $<10^{-4}$         & $0.024^{+1.144}_{-0.024}$                            &  $2.4^{+1.4}_{-0.7}$                 \\
$\log_{10}$sSFR [yr$^{-1}$] &   $<-15$         &   $-12.77^{+1.7}_{-9.28}$                           &  $-7.8^{+0.5}_{-0.3}$                 \\
Av   & $0.02^{+0.05}_{-0.01}$           &  $0.64^{+0.44}_{-0.22}$                             &   $0.88^{+0.20}_{-0.13}$               
\end{tabular}\label{tab:SEDtab}
\end{table*}

\subsubsection{Assessing the Presence of a Radio AGN}

The NEP-TDF field is rich in multi-wavelength data sources. We conduct a search in archival works to assess whether our system is detected in wavelengths beyond the optical and near-infrared in order to further characterise their nature.

The foreground elliptical appears in the catalogue of JWST-Radio crossmatches from \citet{Willner2023}, listed as object ID 182 in the VLA 3GHz catalogue presented in \citet{Hyun2023}. The source has a flux density of $S_{\rm 3GHz} = 74.3\pm2.6$ $\mu$Jy and a compact size of 0.31\arcsec. Assuming it can be attributed to the elliptical galaxy at a redshift of $z=1.258$, this produces a radio luminosity of $\log_{10}(L_{\rm 1.33\,GHz}) = 23.85$~W~Hz$^{-1}$. This luminosity places the source just faintwards of the knee of the AGN radio luminosity function or on the bright-end slope of the galaxy radio luminosity function \citep{Novak2018}. Employing the use of higher resolution radio imaging provided by the Very Large Baseline Array (VLBA, see Saikia et al. In prep), this source is detected with a flux density of $S_{\rm 4.8GHz} = 78\pm10$ $\mu$Jy with an unresolved morphology and centralised in the elliptical galaxy. Its flat radio spectrum (spectral index of approximately 0.1), compact size, and the passive nature of the host indicate that the radio emission originates from an optically thick AGN core. There are no \emph{XMM-Newton} or \emph{NuSTAR} X-ray sources \citep{Zhao2021,Zhao2024} within 20\arcsec\ of NEPJ172238.9+655143.1's position. For the corresponding 1.8 megasecond {\it Chandra} data (Maksym et al. in prep) we use similar methods to \cite{Obrien2024} and \cite{Nabizadeh2024} to place a $3\sigma$ upper flux limit of $\sim3\times10^{-16}$ erg cm$^{-2}$ s$^{-1}$ between 0.5-7.0 keV. {\it Chandra} marginally detects Fe\,K$\alpha$ (3 photons, $\sim9\times10^{-17}$ erg cm$^{-2}$ s$^{-1}$, $\sim1.5\sigma$) using a 100\,eV band at 6.4\,keV (rest frame), supporting evidence from radio that the AGN might be obscured rather than intrinsically X-ray weak.


\subsection{Modelling the Source Galaxy}

\subsubsection{The Redshift of the Source Galaxy}

Using the residual images with the elliptical subtracted, we place a variety of apertures on parts of the ring system, finding broadly consistent EAZY photometric redshifts within the range of $z_s=3.74^{+0.49}_{-0.22}$ and a Balmer break of around 0.7mags located between the F150W and F200W filters. For this galaxy, we assess the use of the default FSPS models in EAZY \citep{Conroy2009,Conroy2010}, as well as the template set from \citet{Larson2022} which features bluer SEDs and stronger emission lines, and obtain similar redshifts. The brightest arc has apparent magnitudes of 25.0 in F444W and 26.1 in F150W, the faintest arc is just over 1 mag fainter and more closely located to residual structure in the lensing elliptical so its photometry is less reliable.

The grism spectroscopy of the background source identifies the continuum but no emission lines. Such a finding is still informative when estimating a redshift and so we conduct a spectro-photometric fit of this source following the same {\tt BAGPIPES} process as above. Despite having no significant spectral features, this information provides tight constraints on the new redshift of $z_s=3.6\pm0.1$, within $\sim1\sigma$ of the original photo-z estimation. This is because with lower redshift solutions, NIRISS would probe and detect H$\beta$ and [OIII] while a higher redshift would place the Balmer break within the F200W bandpass.

\subsubsection{The Gravitational Lens and Magnification Factors}

\begin{figure*}
    \centering
    \includegraphics[width=0.9\textwidth]{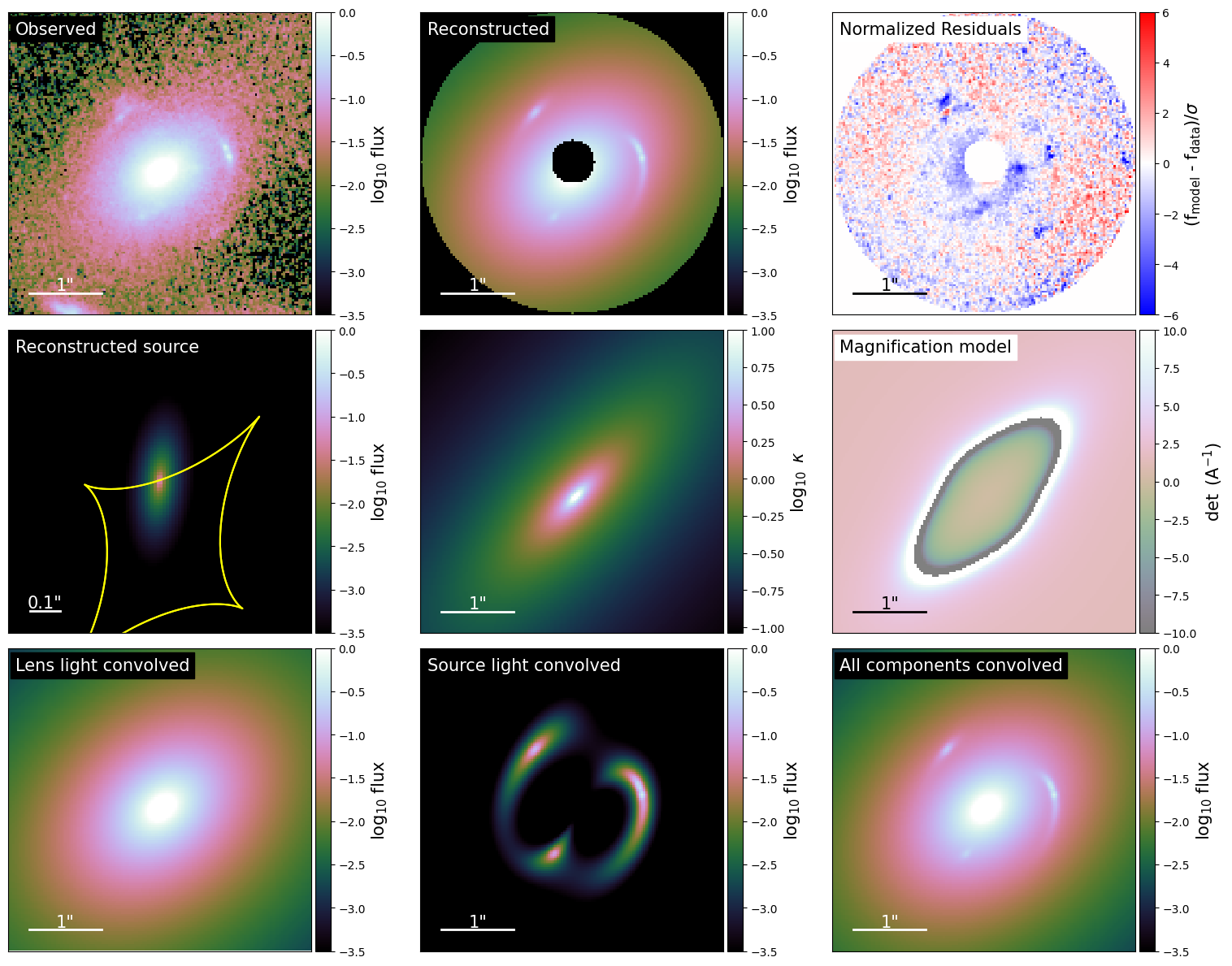}
    \caption{The results of the lens modelling process with {\tt lenstronomy}. The top-left panel shows the original input imaging from JWST F150W. The top-centre panel shows the reconstructed model. The top-right panel shows the residual image when the model is subtracted from the data. The centre-left panel shows the source plane. The central panel shows the intrinsic light profile of the elliptical lensing galaxy. The centre-right panel shows the magnification map. The bottom-left panel shows the elliptical model convolved with the PSF. The bottom-centre panel shows the lensed source, with shear, convolved with the PSF. The bottom-right panel again shows the final reconstructed image.}
    \label{fig:lensmodel}
\end{figure*}

With the redshifts of the lens and source galaxies now tightly constrained, we proceed to model the gravitational lens with the use of the {\tt lenstronomy} code \citep[][]{Birrer2018,Birrer2021}. We model the deflector as a Singular Isothermal Ellipsoid (SIE) mass distribution with external shear, and the deflector and background source light as an elliptical Sérsic profile \citep{Sersic_profile}.
The total mass distribution within the Einstein radius of early-type galaxies is well approximated by an isothermal distribution, such as the SIE (e.g, \citealt{Gavazzi_SLACS_2007}, \citealt{Koopmans_bulge_halo_conspiracy}, \citealt{Lapi_2012}, \citealt{SLACS_debiased}). 
The external shear component accounts for additional shear introduced by line-of-sight perturbers and compensate for the simplicity in the angular structure of the main deflector model (e.g. \citealt{EtheringtonExtShear}).
In the fitting process, we introduce an appropriately sized annular mask to avoid large residuals in the centre of the lens light profile (within $\approx 0.3$", where the Sérsic profile is less adequate in reproducing the observations), and block out the light from the line-of-sight galaxies in the outer region.
We sampled from the posterior probability distribution function (PDF) of the model parameters with the Markov chain Monte Carlo (MCMC) affine invariant method \citep{Goodman_Weare} using {\tt emcee} \citep{emcee}. To achieve faster convergence in the MCMC sampling, we choose the initial points of the Markov-chain from the best-fitting parameters obtained with the particle swarm optimization (PSO) method \citep{ParticleSwarmOptimisation}. The full results from this process are presented in Table \ref{table:lens_params} and Figure \ref{fig:lensmodel}.


We ray-trace the light within elliptical apertures, positioned to enclose the surface brightness of each arc, to calculate magnification factors. For the two brightest arcs (Image 1 and Image 2 in Figure \ref{fig:Field}), we find magnification factors of $\mu\sim8.4$. The two fainter images have magnifications of $\mu=5.04$ (Image 3) for the north western image and $\mu=2.62$ for the southern image (Image 4). When demagnifying aperture photometry of these images, we obtain consistent unlensed fluxes within 0.2 mags, with the exception of the two bluest bands  F606W and F090W (which are both the shallowest bands and the faintest for the source) which are consistent within 0.7mags. When utilising photometry from the source, we use the mean average fluxes of the four demagnified images, these are expressed as the following magnitudes [29.2,28.5,28.3,28.2,27.5,27.5,27.6,27.4,27.3] across the 9 photometric bands used from blue (F606W) to red (F444W).

\begin{table}
    \caption{Best fit lens and source component parameters. The singular isothermal ellipsoid (SIE) component is characterized by the Einstein radius $\theta_E$, the inclination angle $\phi_{SIE}$, and the axial ratio $q$. The shear is defined by the two components $(\gamma_1, \gamma_2)$. The extended source is modelled with an elliptical $R_{S\acute{e} rsic}$ profile, defined by the half-light radius $R_{S\acute{e} rsic}$, the power-law index $n_{S\acute{e} rsic}$ and the center in the source plane. The center of the lens components is kept fixed at $(0, 0)$.}
    \label{table:lens_params}
    \centering
        \begin{tabular}{ |ccccc| }
         
         \hline
         \multicolumn{5}{|c|}{Strong lensing model parameters} \\
         \hline
         \hline
         \multicolumn{3}{|c|}{SIE lens} &\multicolumn{2}{|c|}{SHEAR lens}\\
         \hline
                $\theta_E$ $["]$     & $\phi_{SIE}$ [deg] & $q$     & $\gamma_1$ & $\gamma_2$ \\
         \hline
                $0.92$               & $   46.4$          & $0.38$  &  $0.043$   & $0.097$    \\
         \hline 
               \multicolumn{5}{|c|}{Sérsic source}\\
         \hline
                $R_{\rm{hl}}$ $["] $  & $n$  & $\phi_{src}$ [deg] & $q$  & center$_{\rm{src}}$\\
         \hline
                0.074       & 3.26 & 84.3               & 0.4  & ($0.064, -0.010$)\\
         \hline 
        \end{tabular}

\end{table}

\subsubsection{The Physical Parameters of the Source Galaxy}

Finally, we conduct a {\tt BAGPIPES} fit with the demagnified photometry for the source galaxy in order to obtain corrected masses and star formation rates. Here, we find that the source is a moderately dusty starburst galaxy with a stellar mass of $\log_{10}(M/M_\odot)=8.17^{+0.18}_{-0.23}$, a specific star formation rate that is elevated above the main sequence and moderate dust levels of $A_V=0.88^{+0.20}_{-0.13}$. Its absolute UV magnitude, when demagnified, is estimated to be $M_{\rm UV}=-17.11^{+0.18}_{-0.12}$. Additional information is provided in Table \ref{tab:SEDtab}.

As a sanity check, we also run both the source and lens galaxies with three additional {\tt BAGPIPES} models to assess if the estimated physical parameters are consistent. Two of these models test different star formation history parametrisations (A delayed exponential and a non-parametric continuity model) and one tests the use of the Binary Population and Spectral Synthesis (BPASS) stellar population models \citep{Eldridge2017,Stanway2018,Byrne} versus the default BC03 set \citep{BC03}. We find our source galaxy has best fit parameters that vary within $1\sigma$ relative to the fiducial {\tt BAGPIPES} run. For the total photometry of the foreground elliptical, the continuity model fits for larger amounts of dust ($A_V=0.26^{+0.10}_{-0.07}$), but all other parameters remain consistent between the different runs. The stellar mass enclosed within the 0\farcs92, Einstein radius accounts for $\sim43\%$ of the total stellar mass of the system using a fiducial Kroupa IMF set-up.

\section{Discussion}

\subsection{The chance of observing such a system}

Within the NEP-TDF field, a total of three robust galaxy-galaxy lensing systems are identified in a by-eye search in [Ferrami et al. In Prep]. This value agrees with the range of predictions for the NEP-TDF PEARLS survey made in \citet{Ferrami2024}, which estimated between 3-8 lensing detections depending on two different estimations of the velocity dispersion function (VDF) of elliptical galaxies \citep{Mason2015,Geng2021}. Based on the \citet{Mason2015} VDF \citep[See figure 11 in][]{Ferrami2024}, the Einstein radius ($\theta_E$) of 0\farcs92 lies at the peak of the expected $\theta_E$ distribution. The redshift of the source and the lens are both the highest of the five lensing systems in the NEP-TDF and are around $1.5\sigma$ higher than the anticipated distribution peak. Assuming the system is a Singular Isothermal Ellipsoid (SIE), the velocity dispersion from that lens model is estimated to be $247_{-5}^{+6}$ km s$^{-1}$ which is again close to the peak of the anticipated distribution. The overall lensing geometry of NEPJ172238.9+655143.1 does not cause any tension with expectations of being found in such a field with this area and depth.

\subsection{Mass-Light Ratio and Dark Matter contributions}

With a total lensing mass of $\sim(4.08 \pm 0.07)\times10^{11}M_\odot$ within $\theta_E$, and a corresponding nominal stellar mass of $\sim1.26\times10^{11}M_\odot$, the total dark matter mass partaking in lensing within the radius $< \theta_E$ is approximately $\sim2.82\times10^{11}M_\odot$, providing a mass-to-light ratio of close to $\sim3$. These masses and mass-to-light ratio are almost identical to those obtained from the COSMOS-Web system in \citet{Shuntov2025} despite our different geometry.

To assess whether the total lensing mass can indeed be attributed at least partially to dark matter, we follow the procedure carried out in \citet{Mercier2024} to estimate the dark matter mass within $\theta_E$. To begin, we estimate the total halo mass of the elliptical galaxy using the total stellar mass to halo mass relation presented in \citet{Shuntov2022}. Providing the total stellar mass of the elliptical galaxy to this relation, we obtain a total halo mass of $\log_{10}(M_H/M_\odot)= 13.83^{+0.71}_{-0.30}$. The stellar mass of the elliptical lensing galaxy places it in a mass regime where dark matter halo masses exponentially increase compared to the stellar mass, leading to increased uncertainties in the value of the total halo mass \citep[See also][]{Behroozi2013,Girelli2020}. We then assume this halo mass is distributed following the Navarro–Frenk–White (NFW) mass distribution \citep{NFW1996}, with its concentration assigned following the redshift evolution of the concentration parameter $c_{200}$ presented in \citet{Dutton2014}, and integrate this along the line of sight using a cylinder of radius 0\farcs92 ($\sim$7.68kpc at $z=1.258$) corresponding to the Einstein radius $\theta_E$. 

The resultant dark matter content within $\theta_E$ is found to be $M_{DM}(<\theta_E)=4.12^{+3.32}_{-0.96}\times10^{11}M_\odot$. This value is around $0.12^{+0.21}_{-0.09}$dex higher than the lensing model requires. This means the two mass estimations (from the lens modelling and scaling the stellar mass to a DM mass) are not in significant tension and a simple NFW form appears to provide a suitable representation of the dark matter profile.


\subsection{The impacts of the IMF on relative mass contributions}

The conclusions of the previous subsection are dependent upon a string of assumptions that begin with modelling the stellar populations within the foreground elliptical galaxy using a Kroupa initial mass function \citep{Kroupa2001}. In this sub-section, we explore to what degree conclusions can change if the IMF is allowed to change form. Studies of elliptical galaxies, and especially at smaller radii, towards their centres, have raised the question on whether the IMF functional form is different and could provide greater numbers of low-mass stars \citep[known colloquially as Bottom-Heavy, see e.g.][]{Conroy2012,Maciata2024,vanDokkum2024}. Here, we briefly explore how the conclusions regarding stellar vs dark contributions to the mass budget can vary under a different IMF.

A simple, bottom-heavy alternative is the Salpeter parameterisation of the IMF \citep{Salpeter1955}. A conversion between stellar masses estimated by the Kroupa IMF and the Salpeter IMF can be estimated with by dividing the Kroupa-based mass by a factor of approximately 0.66 \citep{Madau2014}, increasing the stellar mass by $\sim50\%$. In this case, the stellar mass accounts for $\sim1.91\times10^{11}M_\odot$ and its contribution to the lensing mass, as estimated from the lens model, increases from $\sim31\%$ to $\sim47\%$. 

Continuing through with the stellar to halo mass relation and the NFW model, the $1\sigma$ lower limit of the Dark matter mass would be $\sim5\times10^{11}M_\odot$ and the upper range could exceed $10^{12}M_\odot$ (though the higher stellar mass results in increasing degrees of extrapolation from \citet{Shuntov2022}). The total (DM plus stellar) mass of the system would 
be greater than $\sim6.9\times10^{11}M_\odot$ or $>0.22$dex higher than the lensing model requires, resulting in increased tension between the lens model and the simple NFW DM model.

Finally, we explore the scenario where the stellar mass could entirely explain the lensing mass on its own, or at least a large fraction. Higher redshift studies of passive galaxies have identified higher stellar contributions to dynamical masses when compared to lower redshift or local galaxies \citep[e.g.][]{Mendel2020,Kriek2024,Carnall2024} which reduces the need for significant dark matter contributions

A simple calculation can demonstrate what slope of the IMF would be required for stars to provide the full lensing mass. In this scenario, the total stellar mass under a power-law IMF:  \( \xi(M) \propto M^{\alpha} \) is given by the integral:

\begin{equation}
M_{\text{*}} \propto 
 \int_{M_{\text{min}}}^{M_{\text{max}}} M \cdot \xi(M) \, dM
\end{equation}

For a power-law IMF \( \xi(M) = A M^{\alpha} \), the total mass can be written as:

\begin{equation}
M_{\text{*}} \propto \int_{M_{\text{min}}}^{M_{\text{max}}} M^{1+\alpha} \, dM
\end{equation}

\noindent For the Salpeter IMF (\(\alpha = -2.35\)), this gives a total mass \( M_{\text{Salpeter}} \). To double this mass, we require a new slope \( \alpha' \) such that:

\begin{equation}
M_{*} = 
\int_{M_{\text{min}}}^{M_{\text{max}}} M^{1+\alpha'} \, dM = 2 \times \int_{M_{\text{min}}}^{M_{\text{max}}} M^{1-2.35} \, dM.
\end{equation}

\noindent If we assume a typical stellar mass limits of \( M_{\text{min}} = 0.1 M_{\odot} \), \( M_{\text{max}} = 100 M_{\odot} \), we find that a IMF slope of $\alpha' \approx -1.7$ is needed to account for all the observed lensing mass. Such a slope is still bottom-heavy compared to the likes of the Kroupa IMF in the low-mass regime, but simultaneously more top-heavy. Other works have invoked a simultaneously bottom- and top-heavy IMF to begin connecting the developed, massive elliptical cores in the early Universe with modern day ellipticals. Within \citet{vanDokkum2024}, a so-called `Concordance' or `ski slope' IMF for massive ellipticals is proposed which also enables for simultaneously bottom- and top-heavy IMFs by allowing variable power law slopes between different mass regimes. The derived functional form for the IMF in that study was:

\begin{equation}
  \begin{aligned}
  \alpha = -2.4\pm0.09 & \qquad 0.1M_\odot<M<0.5M_\odot \\
  \alpha = -2.0\pm0.14 & \qquad 0.5M_\odot<M<1M_\odot  \\
  \alpha = -1.85\pm0.11 & \qquad 1M_\odot<M<100M_\odot.  \\
  \end{aligned}
\end{equation}

\noindent When employing the mean values for this IMF form, the stellar mass increases by a factor of 1.77 relative to the constant $\alpha=-2.35$ Salpeter IMF. In this scenario, the stellar mass would account for $\sim83\%$ of the lensing mass when compared against the lensing model.


Ultimately, more information is required to fully constrain the problem, such as a secure spectroscopic redshift for the background source galaxy and dynamical information from a higher resolution spectrum of the foreground elliptical. IMFs which result in increased stellar masses consequently necessitate reduced dark matter contributions in order to recreate the observed gravitational lensing. NEPJ172238.9+655143.1 will be included amongst the target lists of future Binospec observations \citep{Fabricant2019} of the NEP-TDF field, in order to extend spectroscopic coverage of the elliptical galaxy to bluer wavelengths of 0.4-0.95microns, and this system would be a target of interest for future IFU follow-up from both ground- and space-based observatories.

\section{Conclusions}

We present the discovery and initial analysis of a high-redshift strong galaxy-galaxy lens which produces a near-Einstein ring. The system is located within the North Ecliptic Pole Time Domain Field targeted by the JWST/PEARLS and HST/TREASUREHUNT survey programmes. We use HST/ACS, JWST/NIRCam and JWST/NIRISS data which provide a spectro-photometric redshift of $1.258\pm0.005$ for the foreground elliptical galaxy as well as a constrained photo-z of $3.6\pm0.1$ for the background lensed galaxy. From this lensing system we identify the following main conclusions:

\begin{enumerate}
    \item The lensing elliptical galaxy is a very massive and passive system at a redshift of $z=1.258\pm0.005$ with a high stellar mass of $\log(M/M_\odot)=11.48^{+0.06}_{-0.15}$, with a negligibly small star formation rate, UVJ colours consistent with passive classification, and observational evidence of a compact, optically thick, radio AGN.
    \item The lensing model identifies 4 separate images which are magnified by factors of $\mu=2.6-8.4$ in a near-Einstein ring geometry. The lensing mass is $\sim(4.08 \pm 0.07)\times10^{11}M_\odot$, about 3 times higher than the stellar mass estimated within the Einstein radius $\theta_E$ using a standard SED modelling set-up and a Kroupa IMF.
    \item Modelling the SED of the demagnified background source, we find this system to be a moderately dusty but highly star forming galaxy with stellar mass $\log(M/M_\odot)=8.17^{+0.18}_{-0.23}$,  $\log(sSFR_{10Myr})=-7.8^{+0.5}_{-0.3}$, $M_{\rm UV} = -17.11^{+0.18}_{-0.12}$ and $A_V = 0.88^{+0.20}_{-0.13}$.
    \item The redshifts and Einstein radius of the system are found not to be in tension with predictions of galaxy-galaxy lensing yields from a field employing the same depths and survey geometry as the NEP-TDF from \citet{Ferrami2024}.
    \item The estimated non-luminous mass, based on the lensing model, within the Einstein radius is $\sim2.82\times10^{11}M_\odot$. When employing simple assumptions and scaling relations, we obtain a predicted dark matter mass which is $0.12^{+0.21}_{-0.09}$dex higher than the mass required, with the caveat that our lensing system lies at the very high-mass end of the employed relations.
    \item If the IMF of the foreground elliptical galaxy is more bottom-heavy, as suggested by some studies of stellar populations within ellipticals \citep[see][for a review]{Smith2020}, then the stellar mass of the lensing system further increases. The system subsequently requires a smaller dark matter contribution in order to avoid raising tension with the mass required to produce the observed lensing.
\end{enumerate}

Current JWST searches within the field of strong galaxy-galaxy lensing have been limited both in area and have also been driven by chance, by-eye, identification in early imaging. With over 3 years of operations completed, JWST surveys now provide multiple programmes with areas on the scale of 75-250 square arcminutes (e.g. CEERS, NEP-TDF, PRIMER-UDS, PRIMER-COSMOS) and some with even larger areas of 500-2500 square arcminutes (COSMOS-Web, PANORAMIC, BEACON). There is subsequently significant scope for more complete systematic searches for galaxy-galaxy lensing systems with JWST \citep[][Ferrami et al. In Prep]{Nightingale2025} and \emph{Euclid} \citep{Walmsley2025} which will enable more discoveries of higher redshift instances and allow us to study such systems from a population perspective. \emph{Euclid} will provide large samples of elliptical lens galaxies with different Einstein radii and different redshifts which JWST can follow up with greater wavelength coverage and spectroscopic capabilities, opening up opportunities to explore the elliptical galaxy IMF and DM halo concentrations in more detail.

Expanding further, large samples with IFU coverage will enable deeper explorations into the dynamics and stellar populations within the foreground galaxies acting as lenses (e.g. Cycle 3's upcoming programme 5883 will observe the COSMOS-Web system with the NIRSpec IFU). The relative brightness of these galaxies can enable such information to be obtained in 1-10hrs per galaxy-galaxy lensing system.

\section*{Acknowledgements}

We acknowledge support from the ERC Advanced Investigator Grant EPOCHS (788113), as well as two studentships from STFC. This research was supported by the Australian Research Council Centre of Excellence for All Sky Astrophysics in 3 Dimensions (ASTRO 3D), through project number CE170100013. R.A.W., S.H.C., and R.A.J.
acknowledge support from NASA JWST Interdisciplinary Scientist grants NAG5 12460, NNX14AN10G
and 80NSSC18K0200 from GSFC. C.N.A.W  acknowledges funding from the JWST/NIRCam contract to the University of Arizona NAS5-02015.

This work is based on observations made with
the NASA/ESA Hubble Space Telescope (HST)
and NASA/ESA/CSA James Webb Space Telescope
(JWST) obtained from the Mikulski Archive for
Space Telescopes (MAST) at the Space Telescope Science Institute (STScI), which is operated by the Association of Universities for Research in Astronomy,
Inc., under NASA contract NAS 5-03127 for JWST,
and NAS 5–26555 for HST. The observations used in
this work are associated with JWST programme 2738 and raw images are available on the MAST archive. This work is based on observations associated with programs HST-GO-15278, 16252 and 16793 made with the NASA/ESA Hubble Space Telescope. We thank our Program Coordinator, Tricia Royle, for her expert help scheduling this HST program. The authors thank all involved in the construction and operations of the telescopes as well as those who designed and executed these observations, their number are too large to list here and without each of their continued efforts, such work would not be
possible.

The authors thank Anthony Holloway and Sotirios
Sanidas for their providing their expertise in high performance computing and other IT support throughout
this work. 

\section*{Data Availability}

JWST data from PEARLS programme ID 2738 is is publicly available on the MAST portal and accessible at \url{https:doi.org/10.17909/d9b7-0056}. Accompanying HST data from TREASUREHUNT is available from \url{https://doi.org/10.17909/wv13-qc14}



\bibliographystyle{mnras}
\bibliography{mnras_template} 








\bsp	
\label{lastpage}
\end{document}